\documentclass[aps,prl,twocolumn,showpacs,superscriptaddress,groupedaddress]{revtex4}  
\usepackage[caption=false]{subfig} 
\usepackage{graphicx}
\usepackage{epsfig}
\usepackage{epsf}
\usepackage{epstopdf}
\usepackage{hyperref}
\usepackage{ragged2e}
\usepackage{graphicx}  
\usepackage{dcolumn}   
\usepackage{bm}        
\usepackage{amssymb}   

\usepackage{preprintcover}  

\PreprintCoverPaperTitle{Observation of an Excited $B^{\pm}_c$ Meson
  State with the ATLAS Detector}  

\PreprintIdNumber{CERN-PH-EP-2014-137}  

\PreprintCoverAbstract{A search for excited states of the $B^{\pm}_c$
meson is performed using 4.9~fb$^{-1}$ of 7~TeV and 19.2~fb$^{-1}$ of
8~TeV $pp$ collision data collected by the ATLAS experiment at the
LHC. A new state is observed through its hadronic transition to
the ground state, with the latter detected in the decay $B_c^{\pm} \to
J/\psi \pi^{\pm}$. The state appears in the
$m(B_c^{\pm}\pi^{+}\pi^{-})-m(B_c^{\pm})-2m(\pi^{\pm})$ mass
difference distribution with a significance of 5.2 standard
deviations. The mass of the observed state is $6842 \pm 4 \pm 
5$~MeV, where the first error is statistical and the second is 
systematic. The mass and decay of this state are consistent with 
expectations for the second $S\mbox{-}$wave 
state of the $B^{\pm}_c$ meson, $B^{\pm}_c(2S)$.}  

\PreprintJournalName{Physical Review Letters}  

\begin{document}
\title{Observation of an Excited $B^{\pm}_c$ Meson
  State with the ATLAS Detector}
\author{ATLAS Collaboration}
\begin{abstract}
A search for excited states of the $B^{\pm}_c$
meson is performed using 4.9~fb$^{-1}$ of 7~TeV and 19.2~fb$^{-1}$ of
8~TeV $pp$ collision data collected by the ATLAS experiment at the
LHC. A new state is observed through its hadronic transition to
the ground state, with the latter detected in the decay $B_c^{\pm} \to
J/\psi \pi^{\pm}$. The state appears in the
$m(B_c^{\pm}\pi^{+}\pi^{-})-m(B_c^{\pm})-2m(\pi^{\pm})$ mass
difference distribution with a significance of 5.2 standard
deviations. The mass of the observed state is $6842 \pm 4 \pm 
5$~MeV, where the first error is statistical and the second is 
systematic. The mass and decay of this state are consistent with 
expectations for the second $S\mbox{-}$wave 
state of the $B^{\pm}_c$ meson, $B^{\pm}_c(2S)$.
\end{abstract}

\maketitle

The $B^{\pm}_c$ meson was first observed by the CDF
experiment in the semileptonic decay mode \cite{CDF_1998}, and its
hadronic decay mode was observed later by both CDF \cite{CDF_2006} 
and D0 \cite{D0_2008} experiments. The $B^{\pm}_c$ meson has also been
observed by the LHCb experiment in various decay modes \cite{BcLHCb}.
Excited states of the $B^{\pm}_c$ meson have not previously been observed. 
The spectrum and properties of the $B^{\pm}_c$ family are predicted by
non-relativistic potential models, perturbative QCD, and lattice calculations
\cite{strong}. 
Measurements of the ground and excited
states through fully reconstructed 
channels will provide tests of the predictions of these models and ultimately the
opportunity to extract information on the strong interaction potential. 

The second $S\mbox{-}$wave 
state, $B^{\pm}_c(2S)$, is predicted to have a mass in the range
$6835$--$6917$ MeV \cite{strong}. 
The next $S\mbox{-}$wave state,
$B^{\pm}_c(3S)$, 
is predicted to have a mass above the
threshold for decay into a $BD$ meson pair. 
Both the $1S$ and $2S$ states have 
pseudoscalar ($0^-$) and vector ($1^-$) spin states that are predicted to
differ in mass by about 20--50 MeV. Transitions between the spin states
occur through soft photon radiation that escapes identification in ATLAS;
the spin states cannot be separated by this analysis. 

This Letter presents the observation of a new state whose mass is consistent
with predictions for the $B^{\pm}_c(2S)$, the second $S\mbox{-}$wave state of the
$B^{\pm}_c$ meson. The $B^{\pm}_c(2S)$ state is reconstructed in the
decay to the $B^{\pm}_c$ meson and two oppositely charged pions, with the
$B^{\pm}_c$ reconstructed through its decay to $J/\psi \pi^{\pm}$, $J/\psi
\to \mu^+ \mu^-$. 

The ATLAS collaboration at the Large Hadron Collider (LHC) uses a
general-purpose particle detector \cite{ATLAS_2008},
consisting of several subsystems including the inner detector (ID),
the electromagnetic and hadronic calorimeters, and the muon
spectrometer (MS). Muon reconstruction at ATLAS makes use of both the
ID and the MS. 
               
Muons pass through the calorimeters and reach the MS if their
momentum is above approximately 3~GeV. The inner detector features a
three-component tracking system, consisting of two silicon-based
detectors, the pixel detector and the microstrip semiconductor tracker
(SCT), and the transition radiation tracker (TRT). ID tracks are
reconstructed if their transverse momentum, $p_{\mathrm T}$ \cite{pT},
is greater than 400 MeV and the magnitude of their pseudorapidity,
$|\eta|$, is less than 2.5.
Muon candidates are formed from a stand-alone MS
track that is matched to an ID track \cite{ATLAS_Muons}. 
In this analysis the MS is only used to identify muons, while their
momentum is measured using the ID information only.

The trigger system \cite{ATLAS_Trigger}
comprises three levels: the hardware-based Level-1 trigger and the
High-Level Triggers (HLT), consisting of
the Level-2 trigger and the Event Filter (EF). The Level-1 trigger
uses resistive plate chambers (RPC) and thin gap chambers (TGC) to
trigger muons in the pseudorapidity ranges of $|\eta|<1.05$ and
$1.05<|\eta|<2.5$, respectively. One or more regions of interest
identified by the Level-1 muon trigger seed the HLT muon online
reconstruction algorithms, where the responses from both the ID and the
MS are combined. For this analysis the HLT selection for the $J/\psi$
requires two muons, respectively $\mu_1$ and $\mu_2$, originating at a
common vertex, with the invariant mass of the muon pair lying between
2.5~GeV and 4.3~GeV.  
The individual muon $p_{\mathrm T}$ thresholds are
$p_{\mathrm{T}}(\mu_1) > 6$~GeV 
and $p_{\mathrm{T}}(\mu_2) > 4$~GeV. 

This study uses $pp$ collision data collected in the years 2011
($\sqrt{s}=$~7~TeV) and 2012 ($\sqrt{s}=$~8~TeV) by the ATLAS
experiment. 
The datasets used correspond to an integrated luminosity of
4.9~fb$^{-1}$ and 19.2~fb$^{-1}$, respectively. The luminosity estimate
has an uncertainty of 1.8\% in 2011 \cite{ATLAS_lumi_2011} and
2.8\% in 2012. 

To improve the resolution, 
peaks are sought in the distribution of the variable $Q=m(B^{\pm}_c
\pi \pi) - m(B^{\pm}_c) - 2m(\pi^{\pm})$, where $m(B^{\pm}_c)$ is the
offline reconstructed invariant mass of the $B^{\pm}_c$ candidate,
$m(B^{\pm}_c \pi \pi)$ is the invariant mass of the $B^{\pm}_c$
candidate combined with two charged pion candidates, and
$m(\pi^{\pm})$ is the charged pion mass \cite{PDG_2012}.
The $B^{\pm}_c$ candidates are reconstructed through the decay
$B^{\pm}_c \to J/\psi(\mu^+ \mu^-) \pi^{\pm}$. Since hadronic decays
of excited $B$ meson states have no observable displacement from the
$pp$ interaction point, these candidates are then combined with two
charged pion candidates both originating at the corresponding $pp$
interaction point. The mass difference ($Q$) distribution 
is formed and analyzed in the range $0-700$~MeV.  

The $B^{\pm}_c$ selection criteria for the events are optimized
separately for 7~TeV and 8~TeV data using the corresponding Monte
Carlo (MC) samples, produced in the ATLAS simulation framework
\cite{ATLSIM, GEANT4}. Differences between the 7~TeV and 8~TeV
selection criteria are due to the higher $\sqrt{s}$ and the number of
simultaneous $pp$ interactions per bunch crossing hereafter called
pile-up. The increase in the $B^{\pm}_c$ production cross section  
between the 7~TeV and 8~TeV MC samples is approximately 3\%. 

The PYTHIA 6 and 8 MC generators \cite{Pythia6, Pythia8}, tuned for
the LHC conditions \cite{ATLAS_tune}, are used along with a
dedicated PYTHIA extension for $B^{\pm}_c$ meson production, based on
calculations from Ref. \cite{Bc_hadronic_production}. 
The following MC samples are used
to optimize the event selection criteria (charge conjugates
implied): $B_c^+$ decay modes $J/\psi\pi^+$, $J/\psi
K^+$, $J/\psi \rho^+$ ($\rho^+\rightarrow\pi^0\pi^+$),
$J/\psi \mu^+ \nu$, $J/\psi \pi^0\pi^+$, $J/\psi \pi^+\pi^-\pi^+$,
as well as $J/\psi X$ produced inclusively or from $b\bar{b}$. The MC
simulated events are treated in exactly the same way as the collision
data, and the same analysis selections are always applied. 
The exclusive ($B^+_c$) channels are generated with PYTHIA 6, and
inclusive ($J/\psi X$) channels are generated with PYTHIA 8.

This analysis involves several steps. 
First the $J/\psi$ candidates are formed as described
below. Then a pion track candidate from the same vertex is added to
form a $B_c^{\pm}$ meson candidate. Finally,
the $B^{\pm}_c(2S)$ candidates are formed by adding two pion
candidates originating at the primary vertex (PV; the selection of the
primary vertex is defined below).

The $J/\psi$ candidates are reconstructed from pairs of oppositely
charged muons. 
These pairs are fitted to a common vertex with the procedure described in
Ref.~\cite{VkalVrt}. The following requirements are applied to each $J/\psi$
candidate:
\begin{itemize}
\item The $p_{\mathrm T}$ of the higher-$p_{\mathrm T}$ muon candidate
  to be above 6~GeV, and the $p_{\mathrm T}$ of the
  lower-$p_{\mathrm T}$ muon candidate to be above 4~GeV; these are
  approximately the $p_{\mathrm T}$ values for which the trigger is fully efficient.
\item The $J/\psi$ vertex fit chi-square per degree of freedom $\chi^2 / NDOF < 15$ (this selection
  keeps more than 99.9\% of the candidates, and is included only to
  remove spurious dimuon combinations).
\item The invariant mass resolution of the
  $J/\psi$  depends on the direction of the muons, with more
  forward candidates showing poorer resolutions.  The sample is split
  into three categories and selection requirements are
  optimized for each of them: (1) both muons have $|\eta|<1.05$, (2) one
  muon has $|\eta|<1.05$ and the other one has $1.05<|\eta|<2.5$, (3)
  both muons have $1.05<|\eta|<2.5$. The invariant mass
  $m(\mu^+\mu^-)$ is calculated from 
  the track parameters adjusted by the common-vertex fit
  to be in a three standard deviation $\pm 3\sigma$ mass window around
  the world average mass $m(J/\psi)$ \cite{PDG_2012}, where
  the $J/\psi$ reconstructed width $\sigma$ varies in the range
  40--90~MeV depending on muon reconstruction precision in different
  regions of pseudorapidity and on the data-taking period. 
\end{itemize}

Selection requirements for the $B^{\pm}_c$ were
chosen by maximizing $S/\sqrt{ S+B }$, using MC events. 
Here $S$ is the number of signal 
events and $B$ refers to the number of background MC events scaled
according to the production cross section. 
The production cross sections are taken from the MC generators'
predictions. The optimization has been carried out
separately for the two different datasets.

The pion candidate from the $B^{\pm}_c$ is
required to have $p_{\mathrm T}>$~4~GeV. Hit requirements in the pixel
detector and SCT are imposed on all tracks to ensure a reliable impact
parameter reconstruction. 

$B^{\pm}_c$ candidates are reconstructed by fitting two muon tracks
from the $J/\psi$ candidate together with a pion
candidate track to a common vertex. The invariant mass of the two
muons is constrained to
the world average of $m(J/\psi)$. The yields per unit of integrated
luminosity, and the central values of the dimuon invariant mass
distributions are verified to be consistent within each dataset.

\begin{figure}[h!]
    \centering
    \includegraphics[width=\columnwidth]{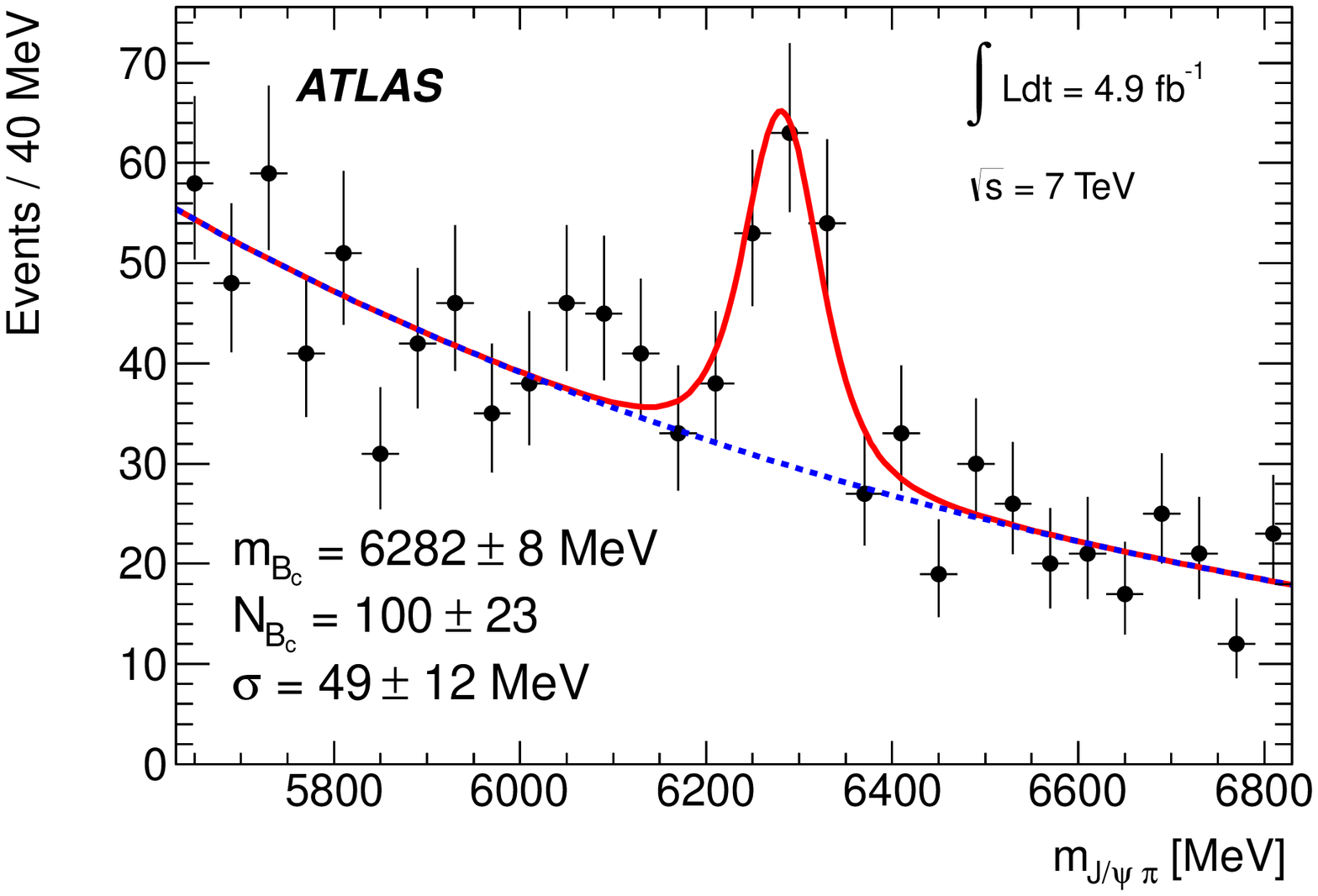}
    \centering
    \includegraphics[width=\columnwidth]{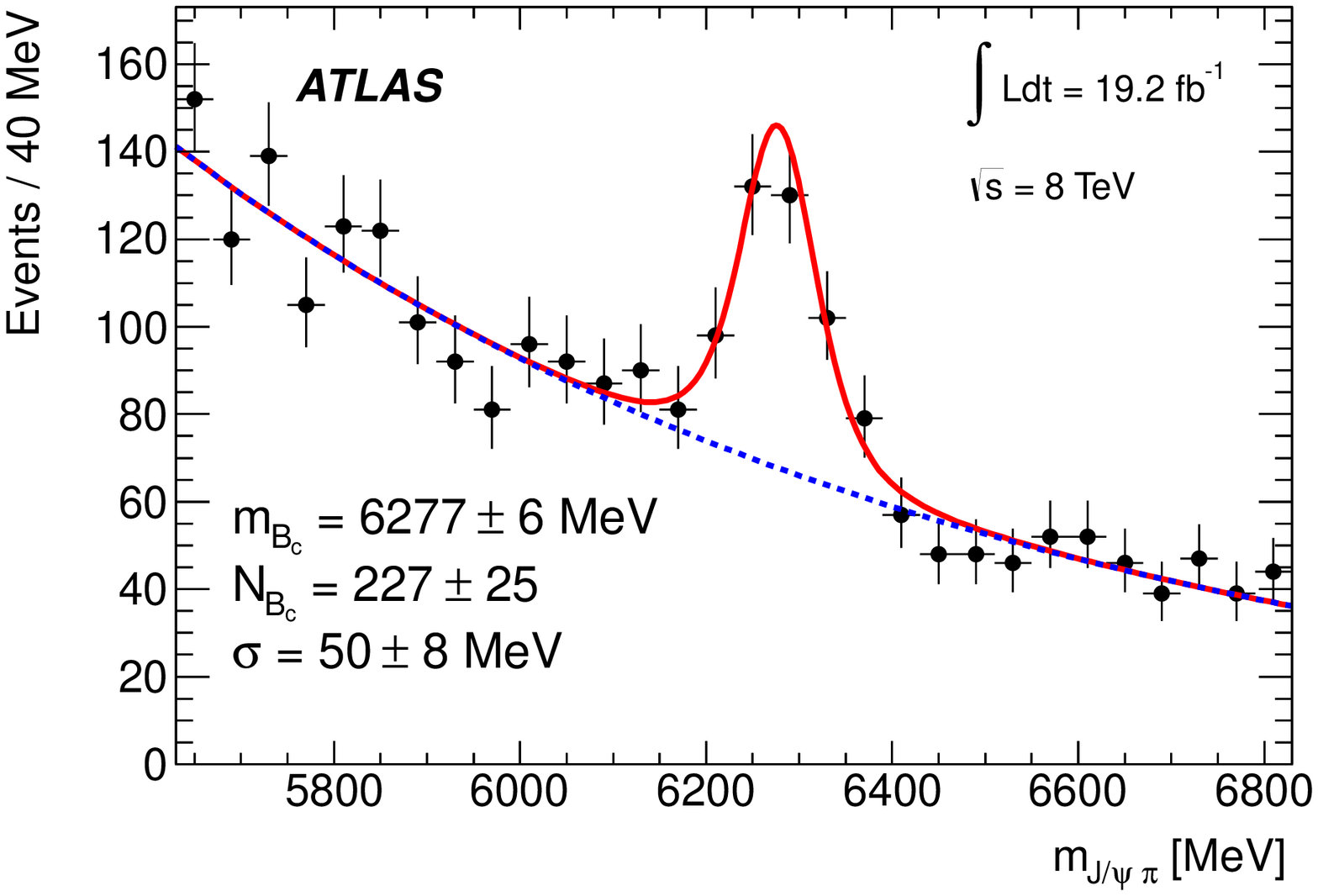}
  \caption{ Invariant mass distributions of the reconstructed
    $B_c^{\pm} \to J/\psi \pi^{\pm}$ candidates in 7~TeV data (top)
    and in 8~TeV data (bottom). The data are represented by the points with error
    bars (statistical only). The solid line is the projection of the results of the
    unbinned maximum likelihood fit to all candidates in the
    mass range 5620--6820 MeV. The dashed line is the projection of
    the background component of the same fit.}
  \label{fig:bc_mass}
\end{figure}

A significant part of the combinatorial background consists of real
$J/\psi$ candidates combined with candidate pions that are
not associated with a $B^{\pm}_c \to J/\psi \pi^{\pm}$ decay. 
This background is reduced by imposing a lower cut on $d_0 /
\sigma (d_0)$; $d_0$ here represents the projection of the pion
track impact parameter relative to the primary vertex onto the
transverse plane,
and $\sigma(d_0)$ is the track-by-track uncertainty on
$d_0$. 
Pion candidates are required to have $d_0 / \sigma
(d_0) > 5$ for 7~TeV data and $ > 4.5$ for 8~TeV data. 
In 7~TeV data the PV is defined as the
collision vertex with the highest summed scalar $p^2_{\mathrm T}$
of its constituent tracks. 
In 8~TeV data -- the data collected with higher pile-up -- 
the PV is identified as the vertex closest in three dimensions to the
reconstructed decay vertex of the $B^{\pm}_c$ candidate. 
All reconstructed
$B^{\pm}_c$ candidates are also required to satisfy the following
selection criteria:
\begin{itemize}
\item the $J/\psi \pi$  vertex fit 
$\chi^2 / NDOF < 2$ for 7~TeV and $ < 1.5$ for 8~TeV data;
\item the $p_{\mathrm T}$ of the $B^{\pm}_c$ candidate to be 
  above 15~GeV for 7~TeV and above 18~GeV for 8~TeV data.
\end{itemize}
Figure \ref{fig:bc_mass} shows the invariant mass distributions for
the $B^{\pm}_c$ candidates for 7~TeV data and 8~TeV data in the mass
range 5620--6820 MeV. Both distributions are fitted separately using
an extended unbinned maximum
likelihood fit, with a Gaussian function modeling the signal and an
exponential modeling the background shape. The various exclusive
$B^{\pm}_c$ decay modes mentioned above are shown to have 
negligible contribution to the background shape.
The results of the fits are summarized in Table \ref{tab:Bfits}. 
The fitted mass values are consistent with the world average for the
$B^{\pm}_c$ mass \cite{PDG_2012}. The signal yield per fb$^{-1}$ of
the $B^{\pm}_c$ is lower in 8~TeV data due to the harder $p_{\mathrm T}$
requirements. The stability of the $B^{\pm}_c$ yield was checked
through its normalization to $B^{\pm} \to J/\psi K^{\pm} $ decays that
were reconstructed with similar requirements.
\begin{center}
\begin{table}[h!]
\centering
  \begin{tabular}{ c | c | c | c }
    \hline
    Data     & Signal events             & Peak mean [MeV]        & Peak width [MeV]  \\ \hline \hline
    7~TeV    & 100$\pm 23$   & 6282$\pm 8$   & 49$\pm 12$ \\ \hline
    8~TeV    & 227$\pm 25$   & 6277$\pm 6$   & 50$\pm 8$ \\ \hline
  \end{tabular}
  \caption{The results of the unbinned maximum likelihood fits of the invariant
    mass distribution of the $B^{\pm}_c$ candidates. Systematic
    uncertainties are not included.}
  \label{tab:Bfits}
\end{table}
\end{center}

The reconstruction of the excited state candidates uses the
$B_c^{\pm}$ ground state candidates within $\pm 3\sigma$ of the
fitted mass value of the corresponding dataset. These candidates are
combined as described below
with two pion candidate tracks associated with the corresponding primary
vertex. The $p_{\mathrm{T}}$ threshold of the pion candidates is
400 MeV. 
No additional selection requirement is applied to the
$B^{\pm}_c(2S)$ pion candidates.
The three tracks from the secondary vertex and the two tracks from the primary
vertex are refitted simultaneously with the
following constraints given by the decay topology: the refitted
triplet of the $B^{\pm}_c$ tracks and the pair of PV pion tracks must
intersect in two separate $B^{\pm}_c$ and $B^{\pm}_c(2S)$ vertices. 
The invariant mass of the refitted muon tracks is constrained
to the $J/\psi$ world average mass, and 
the combined momentum of the refitted $B^{\pm}_c$ tracks must point to
the $B^{\pm}_c(2S)$ vertex. 
When multiple $B^{\pm}_c(2S)$ candidates are found in the same event,
the one with the best $\chi^2$ value
returned by the fitter is kept as an excited state candidate.
Wrong-charge combinations ($B_c^{\pm} \pi^+
\pi^+$ and $B_c^{\pm} \pi^- \pi^-$) are kept separately for comparison
with the combinatorial background shape in the right-charge
combinations (oppositely charged pion pairs).

\begin{figure}[h!]
    \centering
    \includegraphics[width=\columnwidth]{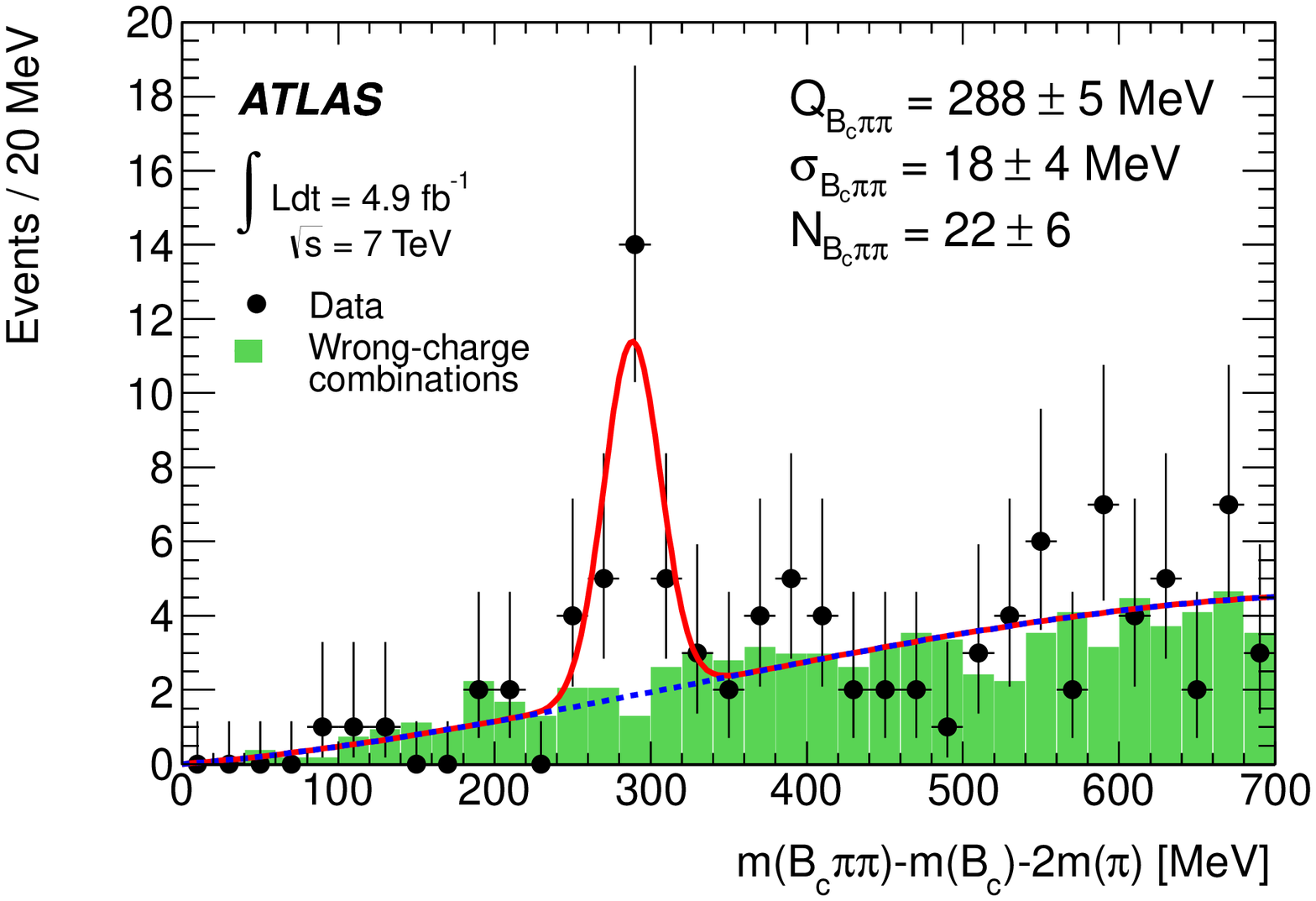}
    \centering
    \includegraphics[width=\columnwidth]{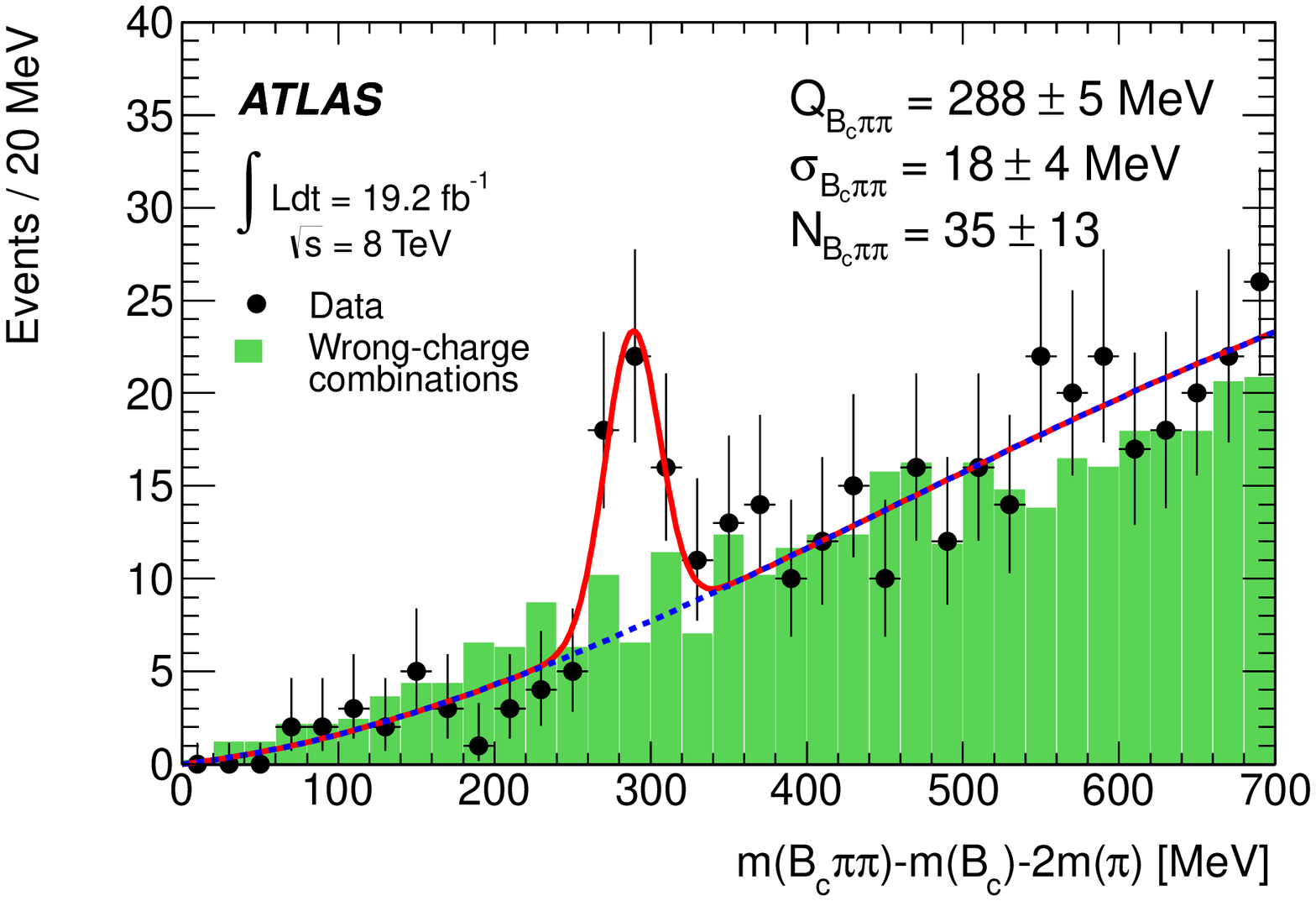}
  \caption{The $Q = m(B_c^{\pm}\pi\pi)-m(B_c^{\pm})-2m(\pi^{\pm})$ 
    distribution for the right-charge combinations (points
    with error bars) and for the same (wrong) pion charge combinations
    (shaded histogram) in 7~TeV data (top) and in 8~TeV data
    (bottom). The wrong-charge combinations are normalized to the same
    yield as the right-charge
    background. The solid line is the projection of the results of the
    unbinned maximum likelihood fit to all candidates in the
    range 0--700 MeV. The dashed line is the projection of the
    background component of the same fit.
}
  \label{fig:bc2S}
\end{figure}

Figure \ref{fig:bc2S} shows the mass difference distribution
$m(B_c^{\pm}\pi \pi)-m(B_c^{\pm})-2m(\pi)$ for the right-charge
combinations $B_c^{\pm} \pi^+\pi^-$ as well as the wrong-charge
combinations. 

A structure is observed in the mass difference distribution. In
order to characterize it, an unbinned maximum likelihood
fit to the right-charge combinations is performed. The fit includes a
third-order polynomial to model the background and a Gaussian function
for the  structure. The background shape resulting from the fit is verified to
be consistent with the wrong-charge combinations (which are not used
to constrain the model in the right-charge fit) by fitting the same
shape to them, with the normalization as the only free parameter.
Alternative models for the signal and the background parameterizations
are studied as sources of systematic uncertainty.
A Breit--Wigner contribution was tested
in convolution with the Gaussian function and was found to be negligible, implying
that the natural width of the structure is small relative to the
detector resolution. The resulting fit parameters (with statistical
uncertainties only), as well as the distributions of the wrong-charge ($B_c^{\pm}
\pi^+\pi^+$, $B_c^{\pm} \pi^-\pi^-$) 
combinations are shown in figure \ref{fig:bc2S}. The wrong-charge
combinations are normalized to the same yield as the right-charge
background.

The relative $B^{\pm}_c(2S)/B^{\pm}_c$ yield ratio is verified to be
statistically consistent between the 7~TeV and 8~TeV data. 

The fit finds the peak at a mass difference ($Q$) value of $288.2\pm 5.1$~MeV in
the 7~TeV data and $288.4\pm 4.8$~MeV in the 8~TeV data. The fit
yields $22 \pm 6$ signal events in the 7~TeV data and $35 \pm 13$ events in the 8~TeV data. The
Gaussian width of the structure is found to be $18.2\pm
3.8$~MeV in the 7~TeV data and $17.6\pm 4.0$~MeV in the 8~TeV data. All
uncertainties mentioned in this paragraph do not include systematic uncertainties.

There are two dominant sources of systematic uncertainty on the position
of the peak. One comes from the uncertainty on the mass of the $B^{\pm}_c$ 
ground state candidate and is largely canceled in the mass difference
distribution. The other involves systematic uncertainties in the fit of
the mass difference distribution itself. The uncertainty on the mass
of the $B_c^{\pm}(2S)$ candidate is dominated by the fitting procedure
and estimated below to be about 3.6~MeV. The contribution
from the uncertainty on the pion momentum scale to the
$B_c^{\pm}$ mass is 1.2~MeV. The residual uncertainty from the $B^{\pm}_c$ candidate
mass in the mass difference distribution, $\delta m_{B^{\pm}_c (2S)} =
\delta m_{B^{\pm}_c} \times (m_{B^{\pm}_c})/m_{B^{\pm}_c(2S)}$, where
the $\delta m_{B^{\pm}_c}$ is the world average uncertainty on the
${B^{\pm}_c}$ mass \cite{PDG_2012}, is about 1.7 MeV. The systematic
uncertainty on the mass difference introduced by the fitting procedure
is estimated by:

\begin{itemize}
\item varying the background model. An exponential threshold function
  ($f(Q) \backsim Q^a e^{-bQ}$, where $a$ and $b$ are free parameters) and
  second- and fourth-order polynomials were considered as
  alternatives, resulting in a 3.4 MeV systematic uncertainty;
\item varying the fit mass range from 0--700 MeV to
  0--1500 MeV, results in a 1.2 MeV contribution to the systematic uncertainty;
\item using different models for the signal. A single Breit--Wigner function, a
  Breit--Wigner function convolved with a Gaussian function, and a
  double Gaussian function were considered. This results in a negligible systematic
  uncertainty, compared to the above two;
\end{itemize}

In each case the largest difference between any of the variations
mentioned and the default fit model is used as the systematic
uncertainty. The values are calculated as the weighted mean of the 
7~and 8~TeV mass values. 

An additional systematic uncertainty of 2~MeV is obtained from the
study of the mass bias in the selection of the candidate with the best
$\chi^2$ of the vertex fit.

The various sources of systematic uncertainty are treated as
uncorrelated. The total averaged systematic uncertainty propagated to
the mass value of the new structure is approximately $4.1$ MeV. 

The significance of the new structure is evaluated with
pseudo-experiments. A large number of background-only mass difference
distributions are generated. Parameters of the generation are taken
from the fit with their uncertainties to account for systematic
effects. The background shape is scaled to the observed number of
events. The mean mass value of the signal contribution is left free to
vary within the theoretically motivated range (6835--6917~MeV) to
evaluate the ``look-elsewhere effect'' \cite{LEE}. The significance is calculated
as the fraction of the pseudo-experiments in which the difference of the
logarithms of fit likelihoods $\Delta$ln$L$ with and without
signal is larger than in the data. In terms of standard deviations the
significance of the observation is 3.7$\sigma$ in the 7~TeV data and
4.5$\sigma$ in the 8~TeV data. 
For the combined 7~TeV and 8~TeV dataset the total significance of the
observation is found to be 5.2$\sigma$. The local
significance of the observation, obtained by fixing the mean value of
the signal component, is 5.4$\sigma$.

In conclusion, the distribution of the mass difference $Q = m(B_c^{\pm} \pi^+ \pi^-) -
m(B_c^{\pm}) - 2m(\pi^{\pm})$ for events with the $B_c^{\pm}$ meson
reconstructed in its decay to $J/\psi \pi^{\pm}$ has been investigated in $pp$
collisions at the LHC using the ATLAS detector. The analysis is based
on an integrated luminosity of 4.9~(19.2)~fb$^{-1}$ of $pp$ collisions
at a center-of-mass energy of 7~(8)~TeV. A new state is
observed at $Q = 288.3 \pm 3.5 \pm 4.1$ MeV (calculated as the error
weighted mean of the 7~TeV and 8~TeV mass values) corresponding to
a mass of $6842 \pm 4 \pm 5$ MeV, where
the first error is statistical and the second is systematic. The
significance of the observation is 5.2$\sigma$ with the
``look elsewhere effect'' taken into account, and the local
significance is 5.4$\sigma$.
Within the uncertainties, the mass of the resonance
corresponding to the observed structure is consistent with the
predicted mass of the $B^{\pm}_c(2S)$ state.

\bigskip

We thank CERN for the very successful operation of the LHC, as well as
the support staff from our institutions, without whom ATLAS could not
be operated efficiently.

We acknowledge the support of ANPCyT, Argentina; YerPhI, Armenia; ARC,
Australia; BMWF and FWF, Austria; ANAS, Azerbaijan; SSTC, Belarus;
CNPq and FAPESP, Brazil; NSERC, NRC and CFI, Canada; CERN; CONICYT,
Chile; CAS, MOST and NSFC, China; COLCIENCIAS, Colombia; MSMT CR, MPO
CR and VSC CR, Czech Republic; DNRF, DNSRC and Lundbeck Foundation,
Denmark; EPLANET, ERC and NSRF, European Union; IN2P3-CNRS,
CEA-DSM/IRFU, France; GNSF, Georgia; BMBF, DFG, HGF, MPG and AvH
Foundation, Germany; GSRT and NSRF, Greece; ISF, MINERVA, GIF, I-CORE
and Benoziyo Center, Israel; INFN, Italy; MEXT and JSPS, Japan; CNRST,
Morocco; FOM and NWO, Netherlands; BRF and RCN, Norway; MNiSW and NCN,
Poland; GRICES and FCT, Portugal; MNE/IFA, Romania; MES of Russia and
ROSATOM, Russian Federation; JINR; MSTD, Serbia; MSSR, Slovakia; ARRS
and MIZ\v{S}, Slovenia; DST/NRF, South Africa; MINECO, Spain; SRC and
Wallenberg Foundation, Sweden; SER, SNSF and Cantons of Bern and
Geneva, Switzerland; NSC, Taiwan; TAEK, Turkey; STFC, the Royal
Society and Leverhulme Trust, United Kingdom; DOE and NSF, United
States of America.

The crucial computing support from all WLCG partners is acknowledged
gratefully, in particular from CERN and the ATLAS Tier-1 facilities at
TRIUMF (Canada), NDGF (Denmark, Norway, Sweden), CC-IN2P3 (France),
KIT/GridKA (Germany), INFN-CNAF (Italy), NL-T1 (Netherlands), PIC
(Spain), ASGC (Taiwan), RAL (UK) and BNL (USA) and in the Tier-2
facilities worldwide.

\onecolumngrid
\clearpage 
\begin{flushleft}
{\Large The ATLAS Collaboration}

\bigskip

G.~Aad$^{\rm 84}$,
B.~Abbott$^{\rm 112}$,
J.~Abdallah$^{\rm 152}$,
S.~Abdel~Khalek$^{\rm 116}$,
O.~Abdinov$^{\rm 11}$,
R.~Aben$^{\rm 106}$,
B.~Abi$^{\rm 113}$,
M.~Abolins$^{\rm 89}$,
O.S.~AbouZeid$^{\rm 159}$,
H.~Abramowicz$^{\rm 154}$
% ...
\bigskip
\\
$^{1}$ Department of Physics, University of Adelaide, Adelaide, Australia\\
% ...
$^{11}$ Institute of Physics, Azerbaijan Academy of Sciences, Baku, Azerbaijan\\
% ...
$^{84}$ CPPM, Aix-Marseille Universit{\'e} and CNRS/IN2P3, Marseille, France\\
% ...
$^{89}$ Department of Physics and Astronomy, Michigan State University, East Lansing MI, United States of America\\
% ...
$^{106}$ Nikhef National Institute for Subatomic Physics and University of Amsterdam, Amsterdam, Netherlands\\
% ...
$^{112}$ Homer L. Dodge Department of Physics and Astronomy, University of Oklahoma, Norman OK, United States of America\\
$^{113}$ Department of Physics, Oklahoma State University, Stillwater OK, United States of America\\
% ...
$^{116}$ LAL, Universit{\'e} Paris-Sud and CNRS/IN2P3, Orsay, France\\
% ...
$^{152}$ Institute of Physics, Academia Sinica, Taipei, Taiwan\\
% ...
$^{154}$ Raymond and Beverly Sackler School of Physics and Astronomy, Tel Aviv University, Tel Aviv, Israel\\
% ...
$^{159}$ Department of Physics, University of Toronto, Toronto ON, Canada\\
% ...
\end{flushleft}



\begin{thebibliography}{99}
\bibitem{pT}
ATLAS uses a right-handed coordinate system with its
origin at the nominal interaction point (IP) in the center of the
detector and the $z$-axis along the beam line. The $x$-axis points from
the IP to the center of the LHC ring, and the $y$-axis points
upward. Polar coordinates ($r$, $\phi$) are used in the transverse ($x$, $y$)
plane, $\phi$ being the azimuthal angle around the beam pipe. The
pseudorapidity is defined in terms of the polar angle $\theta$ as
$\eta = -$ln~tan($\theta/2$).
\end{thebibliography}
\end{document}